\documentclass[letterpaper]{jpconf}
\usepackage{graphicx}
\begin{document}
\title{Finite-frequency dynamics of vortex loops at the $^4$He superfluid phase transition}

\author{Gary A. Williams}

\address{Department of Physics \& Astronomy, University of California,
Los Angeles, CA 90095}

\ead{gaw@ucla.edu}

\begin{abstract}
The finite-frequency dynamics of the $^4$He superfluid phase transition can be 
formulated in terms of the response of thermally excited vortex loops to
an oscillating flow field.  The key parameter is the Hausdorff fractal dimension $d_H$
of the loops, which affects the dynamics because the frictional force on a loop is proportional to the total perimeter $P$ of the loop, which varies as $P \sim a^{d_H}$ where $a$ is the loop diameter.  Solving the 3D Fokker-Planck equation for the loop response at frequency $\omega $ yields a superfluid density which varies at $T_{\lambda}$ as $\omega^{1/(d_H -1)}$.  This power-law variation with $\omega$ agrees with the scaling form found by Fisher, Fisher, and Huse, since the dynamic exponent $z$ is
identified as $z = d_H-1$.  Flory scaling for the self-avoiding loops gives a fractal dimension in terms of the space dimension $d$ as $d_H = (d+2)/2$, yielding $z = d/2 = 3/2$ for d = 3, in complete agreement with dynamic scaling. \end{abstract}

\section{Introduction}
The vortex-loop theory of the superfluid phase transition provides a very simple and physical picture of the transition\cite{williams1,shenoy1,shenoy2}.  An applied superflow acts to orient the thermally excited vortex loops, and their backflow currents screen the net superflow, reducing the superfluid density.  As the temperature is raised towards the lambda point, larger and larger loops can be excited due to screening effects of the smaller loops, and for an infinite system at $T_{\lambda}$ a loop of infinite size drives the superfluid density to zero.  This is not only the simplest renormalization-group theory of the transition, it also now appears to be the most accurate:  two recent simulations \cite{sims} are in agreement with the loop prediction for the superfluid exponent $\nu = 0.671688...$, with the simulations  finding $\nu = 0.6717(1)$.

This simplicity also carries over to the dynamics of the transition, where the characteristic phenomenon of "critical slowing down" can also be easily characterized in terms of the vortex-loop dynamics \cite{dynamics}.  At low temperatures only small loops are excited, and with little frictional damping on them they are able to respond to a time-dependent flow field.  As the temperature gets close to the transition, however, the larger loops being excited are subject to much stronger frictional forces, and can no longer keep up with changes in the applied flow, leading to a rapidly diverging superfluid relaxation time.  The key parameter in this slowing of the dynamic response is the loop Hausdorff fractal dimension $d_H$, since the friction is proportional to the total perimeter $P$ of a loop, which varies as $P \sim a^{d_H}$ where \textit{a} is the average loop diameter. 

For the case of an oscillating applied flow at angular frequency $\omega$, the frictional force gives rise to a frequency-dependent diffusion length,  where loops of diameter considerably smaller than this length can screen the applied flow, while the larger loops cannot.  At the critical temperature  the largest loops no longer respond to the oscillating flow, giving rise to a non-zero superfluid density at and above $T_{\lambda}$.  This is a essentially a finite-size broadening of the transition, with the diffusion length now playing the role of the effective box size.  

The finite-frequency dynamics at $T_{\lambda}$ have also been formulated from dynamic scaling theory by Fisher, Fisher, and Huse \cite{fisher}.  We show here that vortex-loop dynamics give precisely the same results, and a dynamic exponent in full agreement with dynamic scaling.

\section{Vortex loop dynamics} 
In the loop theory the renormalized finite-frequency superfluid density $\rho _s (\omega )$ is given in terms of the zero-frequency density $\rho _s (0 )$ by linear-response theory as \cite{dynamics}
\begin{equation}
\left( {\frac{{\rho _s (\omega )}}{\rho }} \right)^{ - 1}  = \int_{a_0 }^\infty  {\frac{\partial }{{\partial a}}\left( {\left( {\frac{{\rho _s (0)}}{\rho }} \right)^{ - 1} } \right)} \,\,g(\omega ,a)\;da
\end{equation}
where $a_0$ is the core diameter, the smallest loop size, and the response function $g(\omega ,a)$ is found by solving the 3D Fokker-Planck equation \cite{fp} for vortex rings of diameter $a$,
\begin{equation}
g(\omega ,a) = \frac{{\left( {\tilde E - 5/2} \right)}}{{\frac{{ - i{\kern 1pt} \omega {\kern 1pt} a_0^2 }}{{D'}}\frac{\pi }{8}\left( {\frac{a}{{a_0 }}} \right)^{d_H  - 1}  + \left( {\tilde E - 5/2} \right)}} 
\end{equation}
where  terms with derivatives of $g$ have been neglected.  Here  $D'$ is to within a constant the vortex-ring diffusion coefficient given by Donnelly \cite{donnelly} for a ring of diameter $a_0$.  $\tilde E$ is given by
\begin{equation}
\tilde E = \pi ^2 K_r \frac{a}{{a_0 }}(\ln (K^{ - \theta } ) + 1)
\end{equation}
where 
$K_r  = \hbar ^2 \rho _s a_0 /m^2 k_B T$ is the dimensionless renormalized superfluid density,
$K = K_r {\kern 1pt} a/a_0$, and Flory scaling \cite{shenoy2} gives\footnote{The use in Ref. \cite{shenoy2} of a 1/$R$ interaction between vortex segments means that their results are strictly only valid at $d$ = 3. By using the more general interaction $R^{-(d-2)}$ the Flory minimization gives the vlaue of  $\theta$ used here.  The fractal dimension however remains unchanged at $D_H = (d+2)/2$.} $\theta = d/((d+2)(d-2)) = 0.6$ in $d = 3$ dimensions.  At the critical point $T = T_{\lambda}$ an asymptotically exact solution of the loop recursion relations \cite{williams1,shenoy1,shenoy2} gives $K_r = K^* a_0/a$ where $K^* = 0.3875$ is the fixed-point value of K.  Since numerical evaluations show that this $1/a$ dependence is already valid at scale lengths of a few $a_0$, the derivative in Eq.\,1 is essentially a constant except for the smallest loops,  and it is easy to see that $\tilde E$ will then also be nearly constant except for the variation of $K$.  However, at $T_{\lambda}$ $K$ increases rapidly from its initial value $K_0c = 0.309$ at $a_0$ to its fixed point value $K^* = 0.3875$ where it remains constant to the largest loop scales.  Hence it is a very good approximation to take $\tilde E \approx 5$ to be a constant.  In that case the integral of Eq.\,1 is trivial, yielding for the frequency dependence of the superfluid density
\begin{equation}
{\mathop{\rm Re}\nolimits} \,\rho _s (\omega ) \sim \omega ^{1/(d_H  - 1)} 
\end{equation}
From further development of Eq.\,1 it was shown in Ref.\cite{dynamics} that the superfluid relaxation time varies as 
\begin{equation}
\tau  \sim \left( {\frac{{a_0 }}{{K_r }}} \right)^{d_H  - 1}  \sim \xi ^{^{d_H  - 1} } 
\end{equation}
where $\xi$ is the correlation length.  From the definition of the dynamic exponent $z$, the loop prediction is then $z = d_H -1$.  Eq.\,4 is then
\begin{equation}
{\mathop{\rm Re}\nolimits} \,\rho _s (\omega ) \sim \omega ^{1/z} 
\end{equation}
which is the main result of this paper, since it is exactly the $d = 3$ scaling prediction of Fisher, Fisher, and Huse \cite{fisher}.  The imaginary part, the dissipation, also has a universal dependence, which we express in terms of the phase angle
\begin{equation}
\phi  = \tan ^{ - 1} \left( {\frac{{{\mathop{\rm Im}\nolimits} \,\rho _s (\omega )}}{{{\mathop{\rm Re}\nolimits} \,\,\rho _s (\omega )}}} \right) = \frac{\pi }{{2z}}
\end{equation}
which is also the same as in Ref.\cite{fisher}.  The value of $z$ can be found from the Flory-scaling result \cite{shenoy2} for the fractal dimension,  $d_H = (d+2)/2$, giving then $z = d_H-1 = d/2 = 3/2$, showing that the loop dynamics is in complete agreement with dynamic scaling \cite{halperin}.  Numerical simulations of the XY model have now verified this value of $z$ to high accuracy \cite{z}, finding z = 1.500(9).

To check the approximations made in evaluating Eqs.\,1 and 2 above, we have numerically integrated these equations, shown in Figs.\,1 and 2.  Only at very low frequencies is there a very slight difference from Eqs.\,6 and 7, since at low frequencies a wider range of loop sizes is involved before $g$ becomes small.
\begin{figure}[h]
\begin{minipage}{14pc}
\includegraphics[width=20pc]{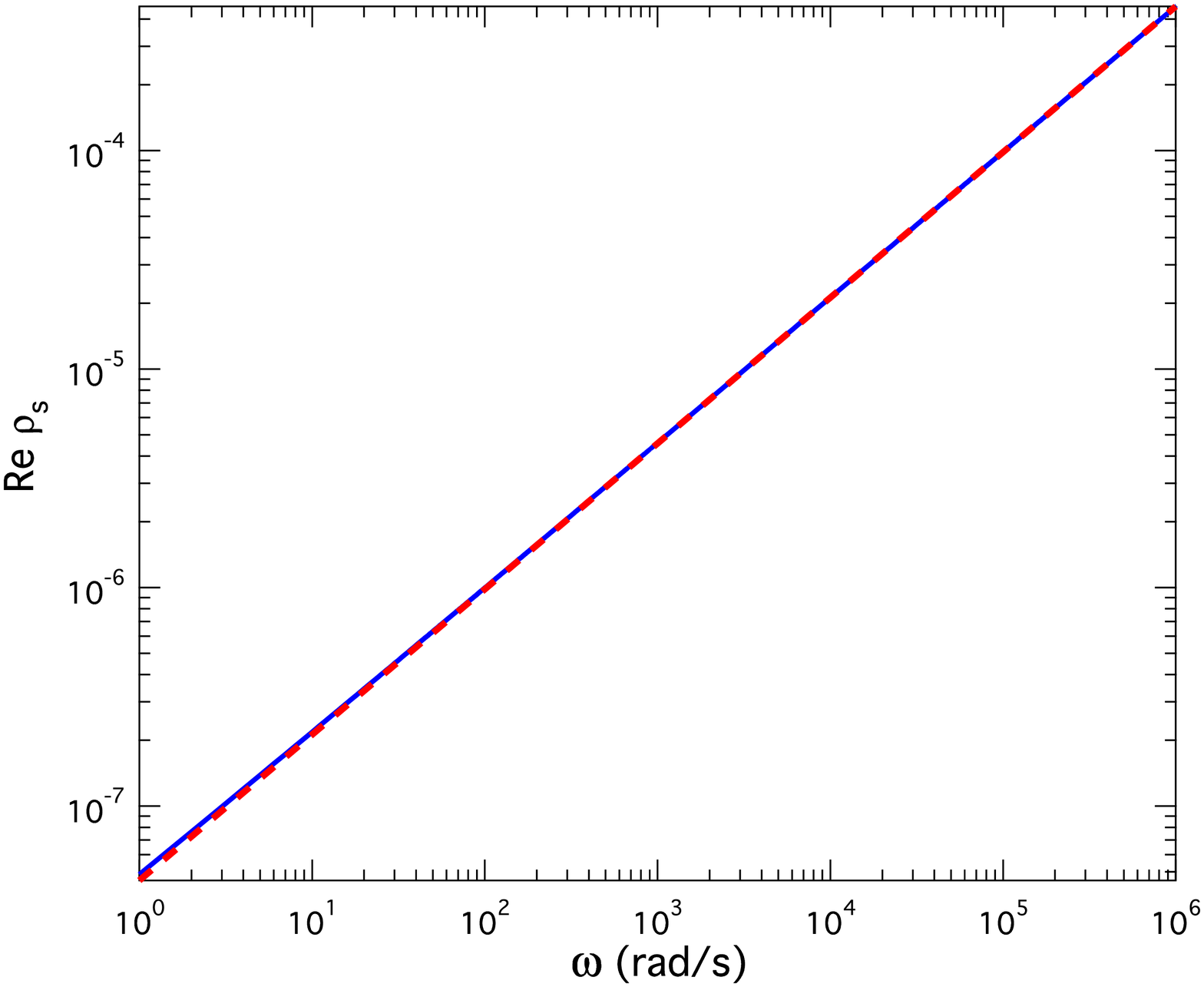}
\caption{\label{label}Full solution of Re $\rho _s (\omega )$ from Eq.\,1 (solid line) compared with $\omega^{2/3}$, Eq.\,6 (dashed line).}
\end{minipage}\hspace{5pc}
\begin{minipage}{14pc}
\includegraphics[width=20pc]{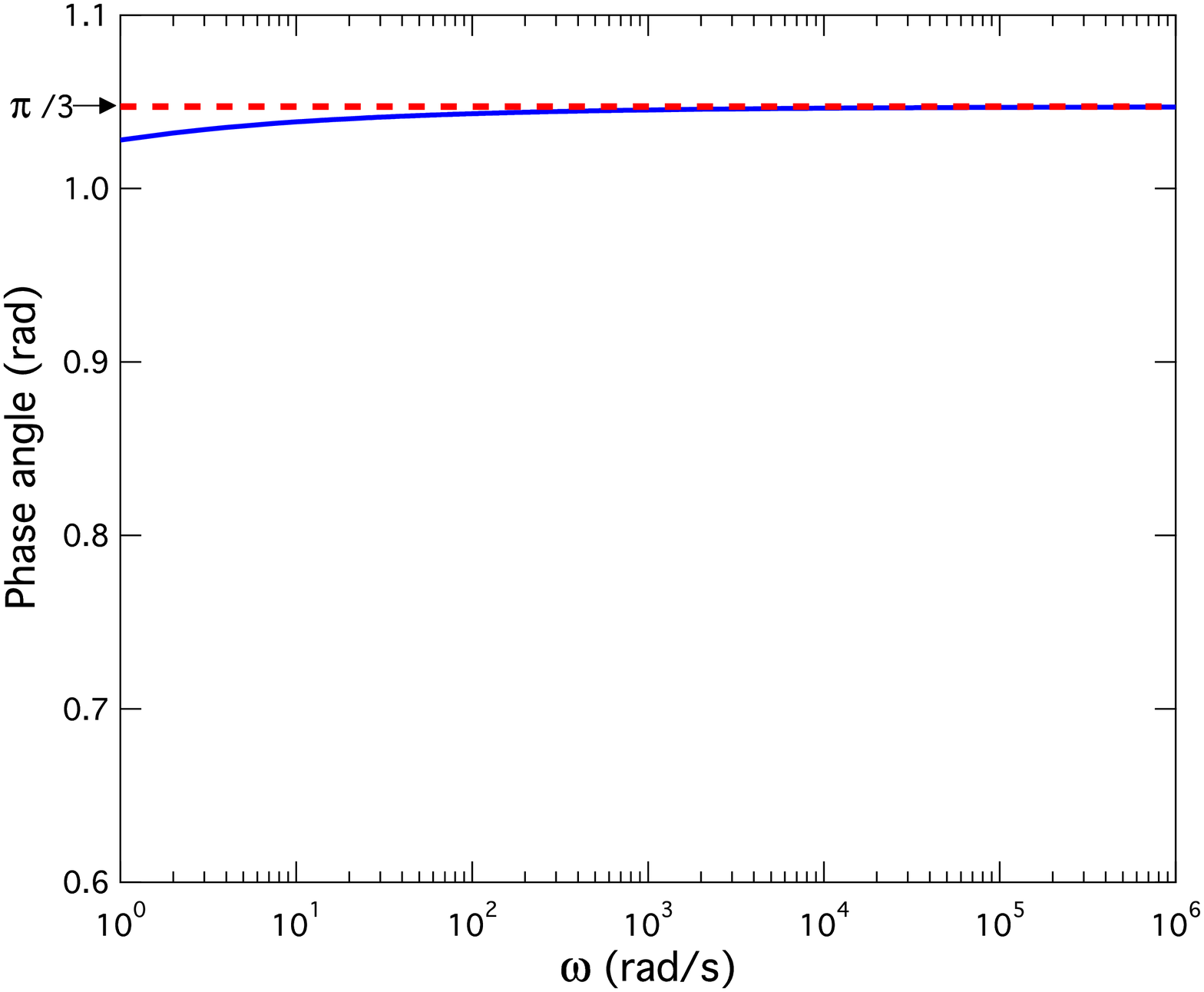}
\caption{\label{label}Full solution for the phase angle (solid line), compared with $\phi = \pi /3$ from Eq.\,7}
\end{minipage} 
\end{figure}

\section{Discussion} 
The agreement with dynamic scaling provides evidence that the Flory scaling result for the loop fractal dimension $d_H = (d+2)/2 = 2.5$ is quite accurate.  Previous estimates \cite{sudbo} of the fractal dimension from XY model simulations have found considerably lower values, however, of 2.287(4) and 2.1693(4).  These authors do not discuss the discrepancy between the two values, which is much greater than the claimed errors.  The fractal dimension is extracted from the computed size distribution of the loops, a procedure which has been noted to be problematic by Olsson \cite{olsson} .  The problem is that  on a lattice two loops that approach each other closer than a lattice spacing (which occurs with some frequency since the density is high near the transition temperature) cannot be resolved, so that there could either be two small loops or one large loop.  A 50-50 guess is made in such a case, with the hope that the effect on the size distribution will average to zero.  The simulations by Olsson gave evidence that this may not be the case, that the guess does lead to a distortion of the distribution.  We note that a current simulation in progress not subject to this problem (the "worm" algorithm) is apparently now finding a value for the fractal dimension that is within a few percent of the Flory value of 2.5 (N. Prokof'ev, private communication).

\ack
This work is supported by the US National Science Foundation, Grant DMR 05-48521.  I thank S. Shenoy, N. Prokof'ev, A. Sudbo, and P. Goldbart for useful discussions.

\end{document}